\documentclass[conference]{IEEEtran}
\IEEEoverridecommandlockouts
\usepackage{cite}
\usepackage{amsmath,amssymb,amsfonts}
\usepackage{algorithmic}
\usepackage{graphicx}
\usepackage{textcomp}
\usepackage{array, makecell}
\usepackage{multirow}
\usepackage{wrapfig}
\usepackage{xspace}
\usepackage{xcolor,colortbl}

\def\BibTeX{{\rm B\kern-.05em{\sc i\kern-.025em b}\kern-.08em
    T\kern-.1667em\lower.7ex\hbox{E}\kern-.125emX}}
\begin{document}

\newcommand{\design}{FSL-HDnn\xspace}
\newcommand{\pnet}{PatterNet\xspace}
\newcommand{\hd}{HDC\xspace}

\title{\design : A 5.7 TOPS/W End-to-end Few-shot Learning Classifier Accelerator with  Feature Extraction and Hyperdimensional Computing}\vspace{-0.4cm}

\author{\large Haichao Yang*, Chang Eun Song*, Weihong Xu, Behnam Khaleghi, Uday Mallappa, Monil Shah,\\
    Keming Fan, Mingu Kang, and Tajana Rosing\\
    \normalsize University of California San Diego, La Jolla, CA, USA;
    E-mail: {cesong@ucsd.edu}, *Equal contributions
} 
\maketitle

\begin{abstract}
This paper introduces \design, an energy-efficient accelerator that implements the end-to-end pipeline of feature extraction, classification, and on-chip few-shot learning (FSL) through gradient-free learning techniques in a 40 nm CMOS process. At its core, FSL-HDnn integrates two low-power modules: Weight clustering feature extractor and Hyperdimensional Computing (HDC). Feature extractor utilizes advanced weight clustering and pattern reuse strategies for optimized CNN-based feature extraction. Meanwhile, HDC emerges as a novel approach for lightweight FSL classifier, employing hyperdimensional vectors to improve training accuracy significantly compared to traditional distance-based approaches. This dual-module synergy not only simplifies the learning process by eliminating the need for complex gradients but also dramatically enhances energy efficiency and performance. Specifically, \design achieves an unprecedented energy efficiency of 5.7 TOPS/W for feature extraction and 0.78 TOPS/W for classification and learning phases, achieving improvements of 2.6$\times$ and 6.6$\times$, respectively, over current state-of-the-art CNN and FSL processors.
\end{abstract}


\section{Introduction}
\begin{figure}
    \centering
    \includegraphics[width=1\linewidth]{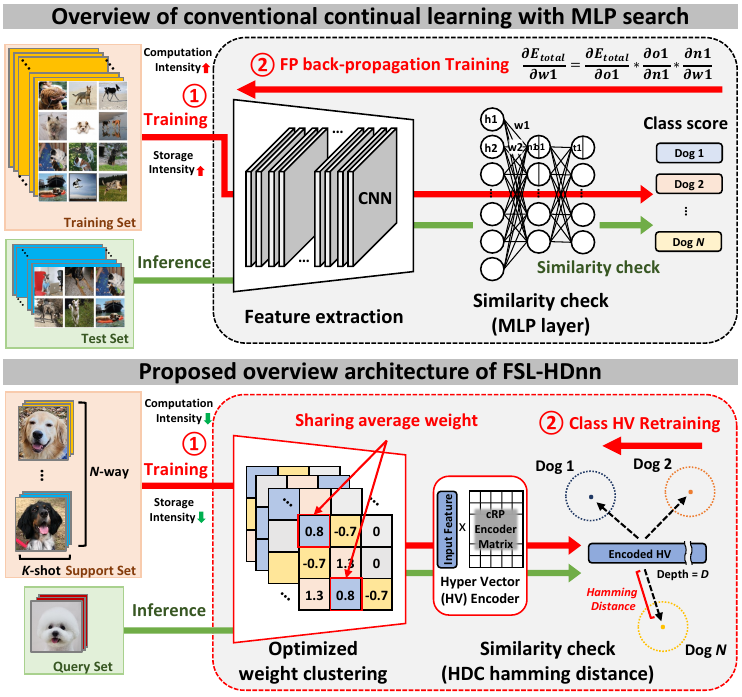}
    \caption{Overview of conventional Few-shot learning pipeline with multilayer perceptron (MLP) search and proposed FSL-HDnn pipeline.}
    \label{fig:overview}
\end{figure}

Continual learning at edge devices is emphasized in many emerging applications to adapt to unseen data and time-varying environments. 
However, on-device learning faces challenges including: 1) learning requires massive training data with limited computation resources in edge device, 2) existing on-chip learning solutions either use back-propagation [5] which is complex and resource-intensive, or simple similarity searches [6] which suffer from low accuracy, 3) feature extraction often incurs high computational costs, such as convolution kernels.
To tackle these challenges, we present a highly efficient end-to-end on-device few-shot learning (FSL) system with hyperdimensional computing (HDC) described in Fig. \ref{fig:overview}. FSL is a machine learning paradigm that quickly adapts to unseen classes with pre-trained weights, requiring fewer than 10 training samples per class.
Although there have been a few existing works on FSL [1, 9] relying on simple similarity checks such as kNN, they suffer from unsatisfactory accuracy. By contrast, The proposed \design leverages the light-weight hyperdimensional computing for the trainable classifier guaranteeing high accuracy, whereas the feature extractor is frozen to boost efficiency.
We demonstrate FSL capability by only retraining the HDC model. \design achieves superior accuracy than simple distance-based FSL (e.g., kNN [6]), while delivering high energy efficiency. 
The feature extractor of \design employs per-filter weight clustering and pattern sharing across filters, which significantly reduces computation complexity.

\section{Proposed Design}
\begin{figure}
    \centering
    \includegraphics[width=1\linewidth]{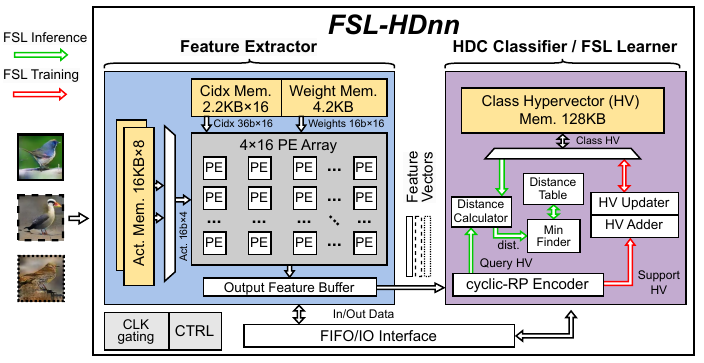}
    \caption{Proposed end-to-end FSL-HDnn Architecture.}
    \label{fig:fsl-overall}
\end{figure}

\design (Fig.~\ref{fig:fsl-overall}) includes 1) feature extractor with weight / index (cidx) / activation memories, and processing elements (PEs) and 2) HDC classifier / FS learner with class Hypervector (HV) memory, HV update module, and similarity checker.  The feature extractor computes  CNN layers with pattern sharing for higher efficiency [2].

\begin{figure}
    \centering
    \includegraphics[width=1\linewidth]{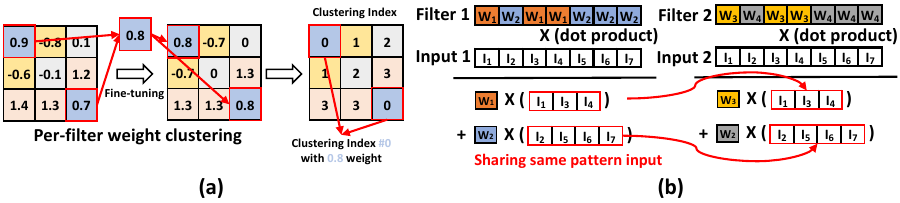}
    \caption{Weight clustering: (a) average weight clustering and index for each weight, (b) accumulated input pixel reuse based on common pattern across filters.}
    \label{fig:pnet_overview}
\end{figure}

HDC classifier performs 1) encoding to convert the feature to HVs, 2) similarity check against HVs to find the closest class HVs from the input for the inference, and 3) FSL by updating the HVs given new data [1]. As shown in Fig.~\ref{fig:pnet_overview}(a), similar weights are clustered into the same average value. Previous studies [7, 8] show that utilizing up to 16 unique weights per filter can achieve accuracy comparable to that of feature extraction processes without implementing weight clustering. This enables weights to be saved as 4-bit indices and indicates a specific pattern of the weight's location in the filter. Also, as shown in Fig.~\ref{fig:pnet_overview}(b), it allows input pixels associated with the same weight to be accumulated together before multiplication. Furthermore, the clustering pattern is shared across filters for different channels so that the accumulated input pixels can be reused by the filters for many output channels. 

\begin{figure}
    \centering
    \includegraphics[width=0.95\linewidth]{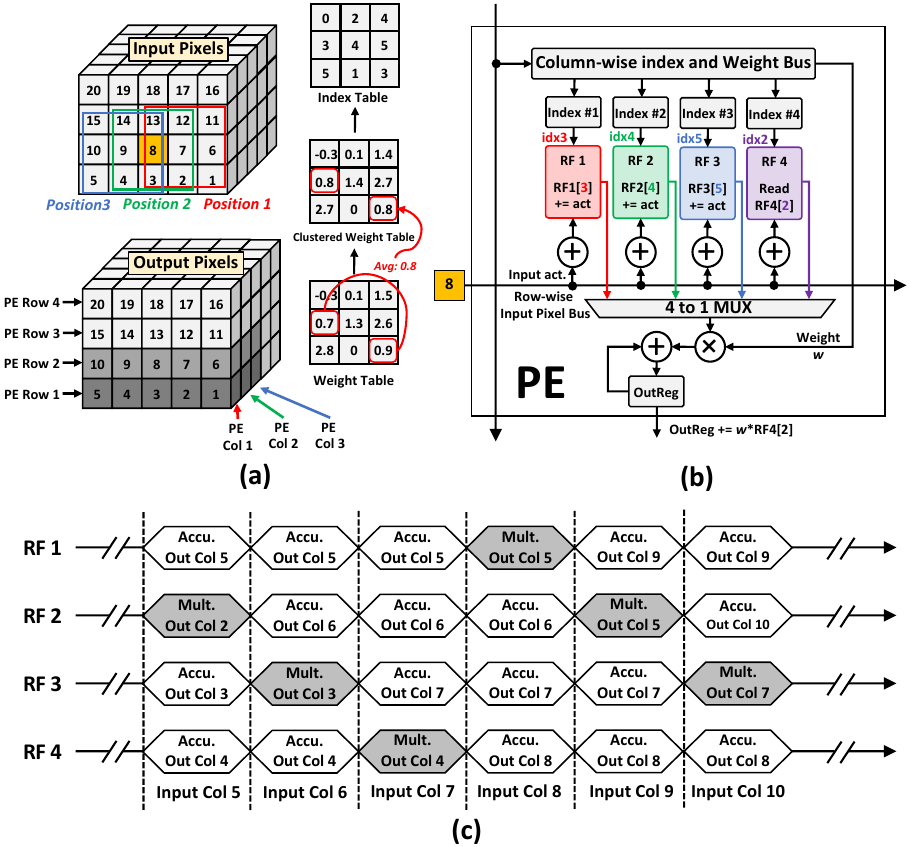}
    \caption{(a) CNN feature extractor with weight clustering, (b) Feature extractor processing element (PE), (c) PE timing diagram.}
    \label{fig:pnet}
\end{figure}

\begin{figure}
    \centering
    \includegraphics[width=0.9\linewidth]{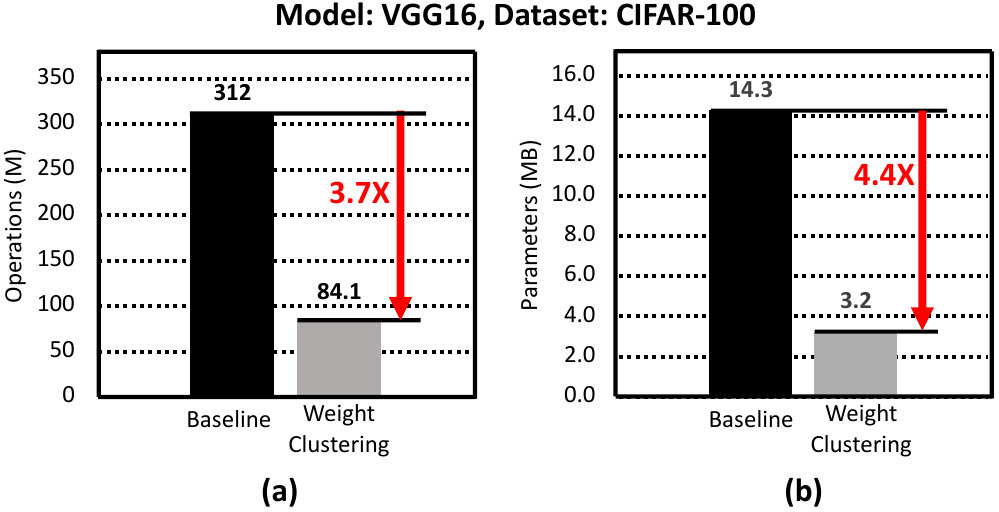}
    \vspace{-0.3cm}
    \caption{Benefits from weight clustering: (a) Operations reduction, (b) Parameters reduction.}\vspace{-0.3cm}
    \label{fig:result3}
\end{figure}

\subsection{Weight Clustering Feature Extractor}
Fig. \ref{fig:pnet} shows the CNN feature extractor to leverage this optimization. The feature extractor (Fig.\ref{fig:fsl-overall} left) contains 64 PEs organized into a 4$\times$16 array. 
PEs on the same row share one input pixel bus, and generate the same output pixel row. PEs on the same column share one index/weight bus, and generate the same set of output pixel channels. The activations associated with the same weight index (i.e., same cluster) are accumulated in the PEs. PEs are optimized for 3$\times$3 convolution kernels. As in Fig. \ref{fig:pnet}(b), each PE contains four Register Files (RFs) that enhance its computational efficiency for convolution operations. Three of these RFs are allocated for accumulating input activations from three separate positions of sliding convolution kernel, allowing parallel processing. For example, in Fig.~\ref{fig:pnet}(a), when the input pixel '8' (colored yellow) is given to the PE, it belongs to three convolutional window positions horizontally neighbored (blue, green, and red). 
The input activation (in this case, pixel 8) is accumulated in three respective RFs based on the index of the group, e.g., for red window position, input '8' goes to the group idx 3 whereas for blue window position, it goes to the group idx 5. The fourth RF is designated for executing multiplication operations with the actual weight values to produce the output pixels. As shown in Fig.~\ref{fig:pnet}(c) timing diagram, this setup ensures that while accumulations for new inputs are underway in three RFs, the fourth can concurrently process multiplications for already accumulated inputs, optimizing the workflow within each PE and enabling more efficient handling of convolution tasks. Due to the proposed efficient feature extracting method, Fig.~\ref{fig:result3}(a) shows that weight clustering achieves 3.7$\times$ and 4.4$\times$ reduction in number of operations and parameters in VGG16, respectively. 

\subsection{\hd Few-shot Learning Module}
\begin{figure}
    \centering
    \includegraphics[width=0.97\linewidth]{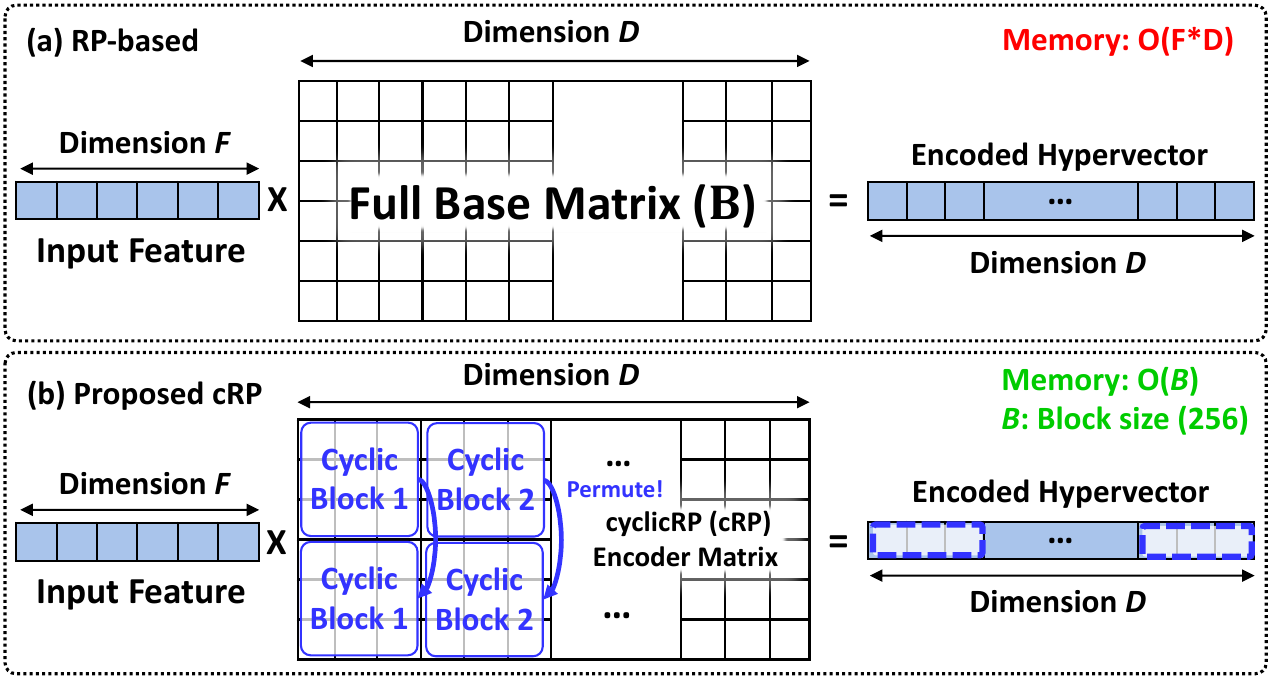}
    \vspace{-0.3cm}
    \caption{(a) Conventional RP-based HDC Encoding (b) Proposed cRP-based HDC Encoding.}\vspace{-0.3cm}
    \label{fig:cpr}
\end{figure}

\begin{figure}[t]
    \centering
    \includegraphics[width=1\linewidth]{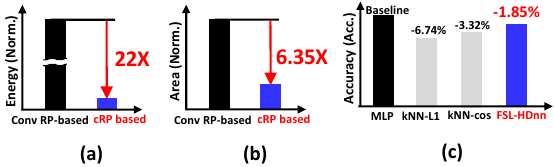}
    \vspace{-0.5cm}
    \caption{Energy, area, and accuracy comparisons: (a) energy efficiency improvement and (b) area efficiency improvement by using cRP-based encoding, (c) accuracy degradation with different distance search methods.}\vspace{-0.3cm}
    \label{fig:fsl_result}
\end{figure}

In Fig.~\ref{fig:cpr}, HDC classifier receives the F-dim feature vector to encode into D-dim HVs for the higher FSL accuracy, where D$\gg $F. The conventional encoding method in Fig.~\ref{fig:cpr}(a) is performed by random projecting (RP) the feature vector on F$\times$D-dim base matrix (\textbf{B}), which is pseudo-random, i.e., randomly generated, but frozen once generated. This encoding method shows promising accuracy, but at the cost of high data volume and access, e.g., N$\times$D for N-class inference. 
We address the overhead by adopting the low-complexity cyclic random projection (cRP) encoder described in Fig.~\ref{fig:cpr}(b), where weights in \textbf{B} are generated on the fly by a cyclic module rather than storing all elements explicitly in buffers. A block of size 256 is loaded into the cRP encoder for each cycle. The cRP encoder reduces 512 - 4096$\times$ memory, 22$\times$ less energy, and 6.35$\times$ less area compared to the original RP encoder (Fig.~\ref{fig:fsl_result}(a) and (b)). 

\begin{figure}
    \centering
    \includegraphics[width=1\linewidth]{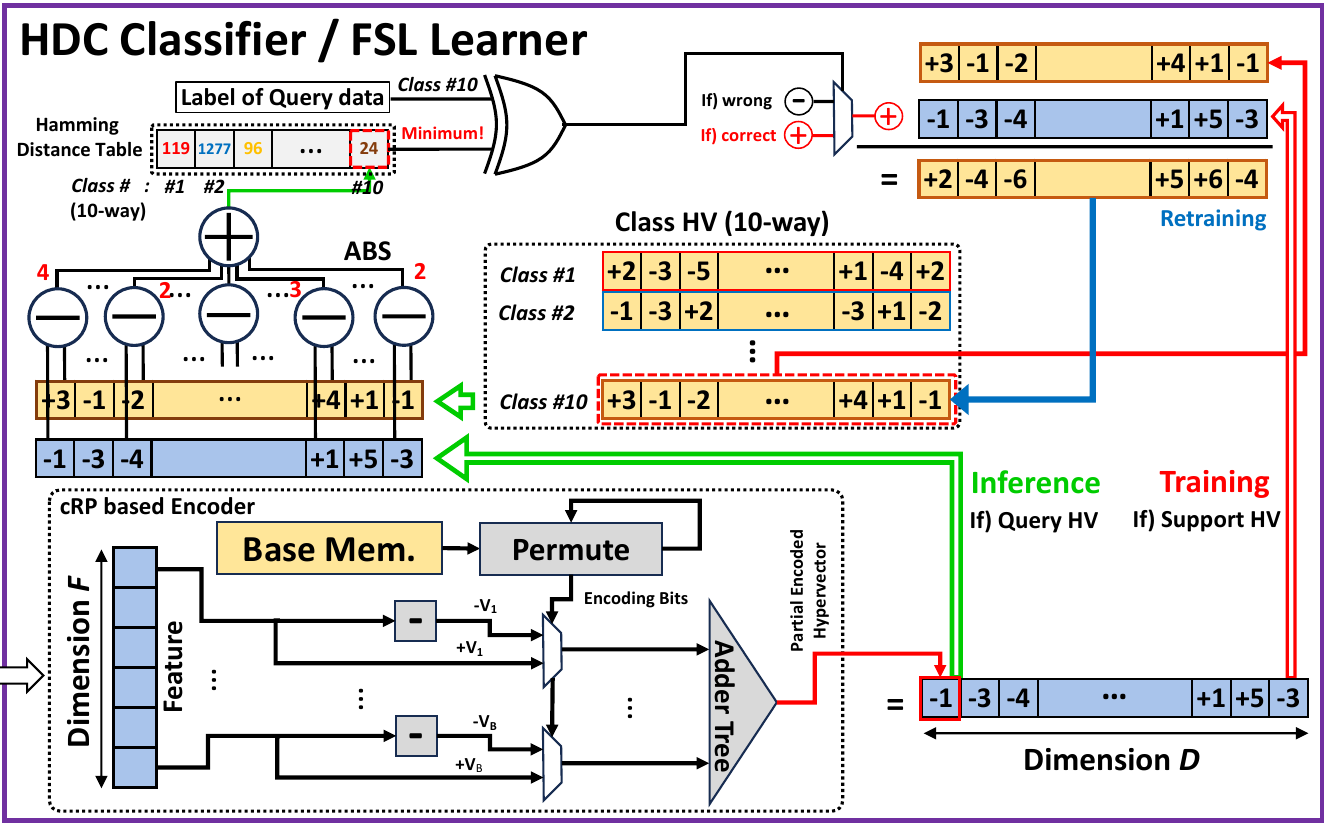}\vspace{-0.3cm}
    \caption{\hd classifier and FS learner with cyclic random projection (cRP) encoding and single-pass FSL.}\vspace{-0.3cm}
    \label{fig:hdc_rp_training}
\end{figure}

HDC classifier (Fig.~\ref{fig:hdc_rp_training}) performs inference by gauging the similarity (Hamming distance) between encoded HV from input and class HVs. HVs are stored in integer format for few-shot training to retain information for future training. During inference, elements of the encoded HV are subtracted from corresponding elements in class HVs. The absolute values of these differences are then accumulated to compute the final Hamming distance. The corresponding class of the HV with a minimum distance from the input HV is the final output of the classifier. The proposed architecture also supports single-pass FSL training with minimal data movement. This is achieved by accumulating the encoded inputs from training data on the chosen class HV if the chosen class by the classifier matches the training label. On the other hand, if the chosen class by the classifier mismatches the training label, the training data will be subtracted from the chosen class HV. All training samples only need to be used once, avoiding repeated data transfer, unlike back-propagation. The proposed architecture has high flexibility allowing the 1-16 bit precisions of HV, 1024 - 8192 for D, 16 - 1024 for F, and the 2-128 classes, which are controllable by the instruction set. Fig.~\ref{fig:fsl_result}(c) depicts that FSL using proposed HDC shows $4.9\%$ FSL accuracy improvements in average over kNN-based designs on various datasets. 


\section{Silicon Measurement and End-to-end Test Results}

\begin{figure}[h]
    \centering
    \includegraphics[width=0.8\linewidth]{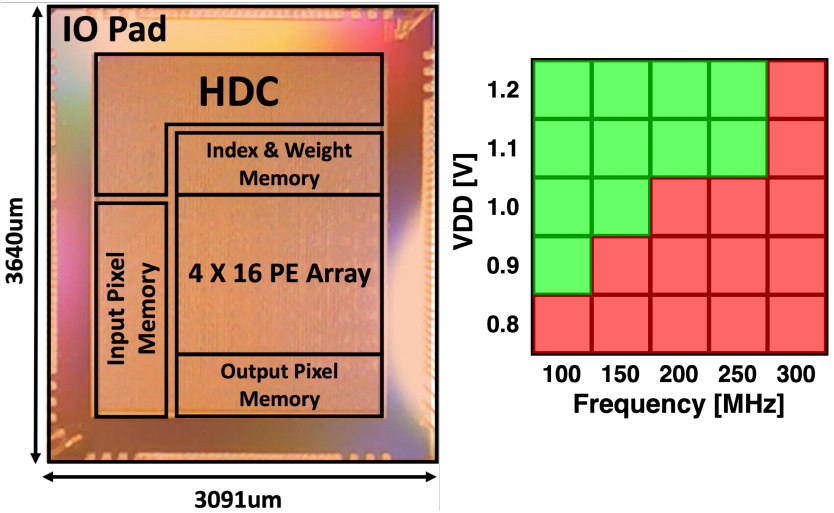}
    \vspace{-0.3cm}
    \caption{Chip micrograph, and shmoo plot.}
    \label{fig:chip_photo_summary}\vspace{-0.4cm}
\end{figure}

 \begin{figure}[ht]
    \centering
    \includegraphics[width=0.95\linewidth]{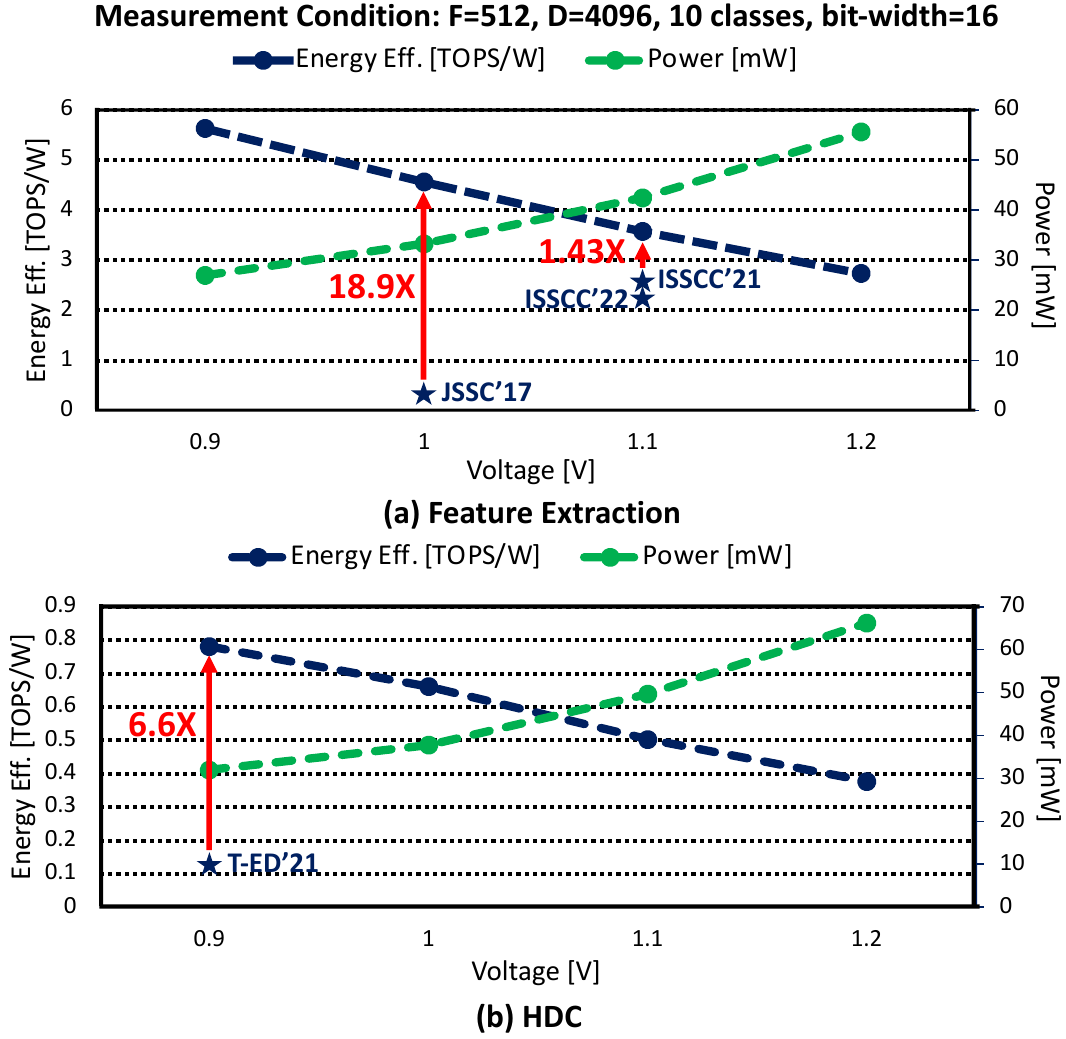}\vspace{-0.3cm}
    \caption{\design measured results for power consumption and energy efficiency with respect to supply voltage for  (a) Feature extraction and (b) HDC.} \vspace{-0.3cm}
    \label{fig:measurement1}
\end{figure}

\begin{figure}
    \centering
    \includegraphics[width=0.95\linewidth]{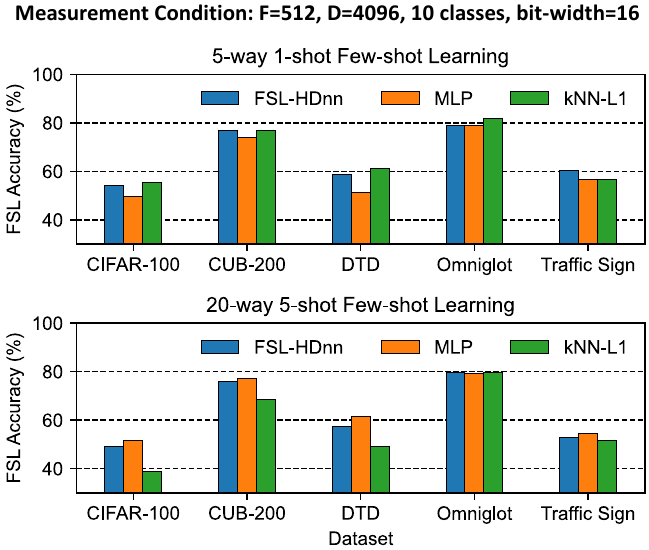} \vspace{-0.3cm}
    \caption{\design accuracy comparison with other techniques for various datasets.} \vspace{-0.3cm}
    \label{fig:measurement2}
\end{figure} 

\begin{figure}
    \centering
    \includegraphics[width=0.95\linewidth]{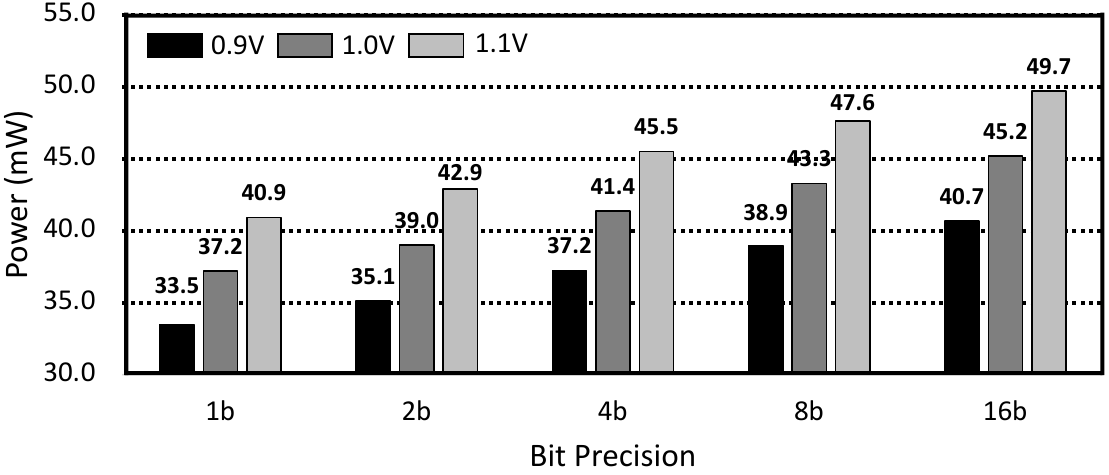} \vspace{-0.3cm}
    \caption{Measured power consumption of HD classifier and FSL blocks based on different bit precision of Class HV for FSL with different supply voltages.} \vspace{-0.3cm}
    \label{fig:result4}\vspace{-0.3cm}
\end{figure}

\design prototype was fabricated in 40~nm CMOS technology with an area of 11.3~mm$^2$. Fig.~\ref{fig:chip_photo_summary} shows the chip micrograph and shmoo plot. We used 349 KB on-chip memory, and deployed BF16 for feature extraction, and INT16 (INT1-16) for HDC FSL training (Inference). The measured results show the operating frequency up to 250~MHz at 1.1~V. Fig.~\ref{fig:measurement1} shows that the power efficiency ranges from 2.8 - 5.7  and 0.38 - 0.78 TOPS/W for the feature extractor and HDC classifier at 0.9 - 1.2~V, respectively, with ultra low-power consumption of 27 and 32~mW at 0.9~V. In Fig.~\ref{fig:measurement2}, the \design accuracy with real benchmark indicates comparable accuracy to the case trained with an MLP-based classifier layer at much lower costs. It also shows much higher accuracy than the trained model based on  kNN-L1  layer. Fig.~\ref{fig:result4} shows measured power behavior with different bit precisions and supply voltages.
Fig.~\ref{fig:comparison} summarizes the comparison with prior FSL and DNN prototypes. Due to the light-weight HDC-based FSL and the feature extraction with pattern sharing, \design achieves $2.6\times$ and $6.6\times$ higher peak TOPS/W than the state-of-the-art CNN and FSL accelerators [3-6], respectively.
Fig.~\ref{fig:chip_summary} shows our chip summarization. 

\colorlet{myGray}{gray!40}

\begin{figure}
    \centering
    \tiny
    \resizebox{\linewidth}{!}{
    \setlength\arrayrulewidth{1pt}
    \begin{tabular}{|>{\columncolor{myGray}}c|c|c|c|c|c|c|}
        \hline
        \rowcolor{myGray}
        & Eyeriss [3] & ISSCC'21 [4] & CHIMERA [5] & SAPIENS [6] & This work \\
        \hline
        Tech. (nm) & 65nm  & 40nm & 40nm & 40nm & 40nm \\
        \hline
        \multirow{2}{*}{ \parbox{1.1cm}{\centering Learning\\ Engine}} & \multirow{2}{*}{\parbox{1.1cm}{\centering CNN-\\BackPropagation}} & \multirow{2}{*}{\parbox{1.1cm}{\centering CNN-\\BackPropagation}} & \multirow{2}{*}{\parbox{1.1cm}{\centering CNN-\\BackPropagation}} & \multirow{2}{*}{\parbox{1.1cm}{\centering kNN-FSL(L1)}} & \multirow{2}{*}{\parbox{1.1cm}{\centering CNN-HDC}} \\
         Engine & & & & & \\
        \hline
         Area (mm$^2$) & 12.25 & 6.25 & 29.2 & 0.2 & 11.3 \\
        \hline
        Freq. (MHz) & 200 & 180 & 200 & 200 & 100 \\
        \hline
        Voltage (V) & 1.0 & 1.1 & 1.1 & - & 0.9 \\
        \hline
        \multirow{2}{*}{\parbox{1.1cm}{\centering Memory \\ Size}} &\multirow{2}{*}{\parbox{1.cm}{\centering 181kB \\ SRAM}}  & \multirow{2}{*}{\parbox{1.cm}{\centering 293kB SRAM}}  & \multirow{2}{*}{\parbox{1.1cm}{\centering 512kB SRAM\\+2MB SRAM}} & \multirow{2}{*}{\parbox{1.cm}{\centering 64kbits RRAM}} & \multirow{2}{*}{\parbox{1.cm}{\centering 349kB SRAM}}\\
        Size & & & & & \\ \hline
        
        Precision & INT16 & FP8  & INT8 & FP32 & BF16 / INT16 \\
        \hline

        Workload & CNN & CNN & CNN & FSL  & \textbf{CNN+FSL}\\
        \hline
        
        \multirow{2}{*}{\parbox{1.1cm}{\centering Peak TOPS \\(CNN)}} & \multirow{2}{*}{\parbox{1.1cm}{\centering 0.067}} & \multirow{2}{*}{\parbox{1.1cm}{\centering 0.567 }}  & \multirow{2}{*}{\parbox{1.1cm}{\centering 0.92}} & \multirow{2}{*}{\parbox{1.1cm}{\centering -}} & \multirow{2}{*}{\parbox{1.1cm}{\centering 0.154}} \\
        (CNN)& & & & & \\ \hline
        
        \multirow{2}{*}{\parbox{1.1cm}{\centering Peak TOPS/W\\ (CNN) }} & \multirow{2}{*}{\parbox{1.1cm}{\centering 0.241 }} & \multirow{2}{*}{\parbox{1.1cm}{\centering 2.5}} & \multirow{2}{*}{\parbox{1.1cm}{\centering 2.2}} & \multirow{2}{*}{\parbox{1.1cm}{\centering - }}  & \multirow{2}{*}{\parbox{1.1cm}{\centering 5.7 }} \\
        (CNN)& & & & & \\ \hline
        
        \multirow{2}{*}{\parbox{1.1cm}{\centering Peak TOPS\\ (FSL)}} & \multirow{2}{*}{\parbox{1.1cm}{\centering -}} & \multirow{2}{*}{\parbox{1.1cm}{\centering -}} & \multirow{2}{*}{\parbox{1.1cm}{\centering -}} & \multirow{2}{*}{\parbox{1.1cm}{\centering 0.0004}} & \multirow{2}{*}{\parbox{1.1cm}{\centering 0.025}} \\
        (FSL)& & & & & \\ \hline
        
        \multirow{2}{*}{\parbox{1.1cm}{\centering Peak TOPS/W \\(FSL)}} & \multirow{2}{*}{\parbox{1.1cm}{\centering -}}  & \multirow{2}{*}{\parbox{1.1cm}{\centering 
        -}} & \multirow{2}{*}{\parbox{1.1cm}{\centering -}} & \multirow{2}{*}{\parbox{1.1cm}{\centering 0.118}} & \multirow{2}{*}{\parbox{1.1cm}{\centering 0.78}} \\ 
        (FSL)& & & & & \\ \hline
        
        \multirow{2}{*}{\parbox{1.1cm}{\centering Power \\(mW)}} & \multirow{2}{*}{\parbox{1.1cm}{\centering 278}} & \multirow{2}{*}{\parbox{1.1cm}{\centering 230}}  & \multirow{2}{*}{\parbox{1.1cm}{\centering 135}} & \multirow{2}{*}{\parbox{1.1cm}{\centering 3.39}} & \multirow{2}{*}{\parbox{1.1cm}{\centering 27 (CNN) / \\ 32 (HDC)}} \\
        (mW) & & & & & \\ \hline

        FSL configs. & No & No & No & No & \textbf{Yes}\\
        \hline
         FSL Feat. dim. & - & - & - & 128 & \textbf{16-1024}\\
        \hline
        On-chip FSL & No & No & No & No & \textbf{Yes}\\
        \hline
    \end{tabular}}
    \caption{Comparison with state-of-art FSL and DNN accelerators.}
    \label{fig:comparison}
\end{figure}

\begin{figure}
    \centering
    \resizebox{0.35\textwidth}{!}{
    \setlength\arrayrulewidth{0.6pt}
    \begin{tabular}{|>{\columncolor{myGray}}c|c|}
        \hline
        Technology & 40 nm \\
        \hline
        Die Size & 11.3 mm$^2$ \\
        \hline
        On-chip Memory & 349 kB \\
        \hline
        Supply Voltage & 0.9 V - 1.2 V \\
        \hline
        Frequency & 100 MHz - 250 MHz \\
        \hline
        Model & CNN + HDC \\
        \hline
        Weight Precision (CNN) & BF16 \\
        \hline
        Weight Precision (HDC FSL) & INT16 \\
        \hline
        Weight Precision (HDC Inference) & INT1-16 \\
        \hline
        FSL Feature Dimension (F) & 16 - 1024 \\
        \hline
        FSL Classes (N) & 2 - 128 \\
        \hline
        HDC Dimension (D) & 1024 - 8192 \\
        \hline
        Power@0.9V, 100 MHz (CNN) & 27 mW \\
        \hline
        Power@0.9V, 100 MHz (HDC) & 32 mW \\
        \hline
        Peak Energy Efficiency (CNN) & 5.7 TOPS/W \\
        \hline
        Peak Energy Efficiency (HDC) & 0.78 TOPS/W \\
        \hline
    \end{tabular}
    \quad}
    \caption{Chip summary.}
    \label{fig:chip_summary}\vspace{-0.3cm}
\end{figure}\vspace{-0.2cm}


\section{Conclusion}
We present \design, a highly efficient 40 nm CMOS accelerator for feature extraction, classification, and on-chip few-shot learning (FSL). Leveraging weight clustering and pattern reuse for energy-efficient CNN-based feature extraction alongside lightweight hyperdimensional computing (HDC) for classification, FSL-HDnn exceeds conventional FSL methods in training accuracy, achieving energy efficiencies of 5.7 TOPS/W for feature extraction and 0.78 TOPS/W for classification. \design demonstrates the feasibility of learning under stringent resource constraints, marking a significant advancement toward on-device learning system at edge. 

\section{Acknowledgements}
This work was supported by TSMC and in part by PRISM and CoCoSys, centers in JUMP 2.0, an SRC program sponsored by DARPA. We would like to thank Carlos Diaz $\&$ Leo Liu for their help with this work, without their suggestions, advice and help during the design and the tapeout, this work would not have been possible.

\vspace{12pt}

\end{document}